\documentclass[journal]{IEEEtran}

\usepackage{amsmath, amssymb, bm, cite, psfrag, mathtools}
\usepackage{rotating}
\usepackage{dblfloatfix}
\usepackage{algorithm,algpseudocode}
\usepackage{array}
\usepackage{supertabular,booktabs}
\usepackage{ragged2e}
\newcolumntype{P}[1]{>{\centering\hspace{0pt}}p{#1}}
\newcolumntype{M}[1]{>{\centering\hspace{0pt}}m{#1}}
\newcolumntype{L}{>{\centering\arraybackslash}m{3cm}}
\usepackage{changepage}
\usepackage[font=small]{caption}
\usepackage{subcaption}
\usepackage{mwe}
\usepackage{color,soul}
\usepackage{booktabs}

\usepackage{capt-of}
\usepackage{bbm}
\usepackage{multirow}
\usepackage[usenames,dvipsnames]{xcolor}
\definecolor{Fgreen}{rgb}{0.13, 0.55, 0.13}

\usepackage{footnote}
\makesavenoteenv{tabular}
\makesavenoteenv{table}
\usepackage{etoolbox}
\usepackage{pbox}
\usepackage[hyphens]{url}
\usepackage{cite}
\usepackage{longtable}
\usepackage{tabu}
\usepackage[margin=0.52in,top=0.4in]{geometry}

\usepackage{fixltx2e}
\usepackage{graphicx,etoolbox}
\usepackage{tabu}

\def\1m{\textrm{1 m}}

\def\MHz{\textrm{MHz}}
\def\kHz{\textrm{kHz}}

\newtoggle{conference}
\togglefalse{conference} 
\usepackage{tikz}
\usetikzlibrary{calc}
\begin{document}
\bibliographystyle{IEEEtran}

\title{A Flexible Wideband Millimeter-Wave Channel Sounder with Local Area and NLOS to LOS Transition Measurements} 
\vspace{-2cm}
\author{
\IEEEauthorblockN{George R. MacCartney, Jr.,~\IEEEmembership{Student Member,~IEEE,}
Hangsong Yan,~\IEEEmembership{Student Member,~IEEE,}\\
Shu Sun,~\IEEEmembership{Student Member,~IEEE,}
and Theodore S. Rappaport,~\IEEEmembership{Fellow,~IEEE}\\
\IEEEauthorblockA{NYU WIRELESS\\
	NYU Tandon School of Engineering, New York University, Brooklyn, NY 11201}\vspace{-0.8cm}
}

\thanks{This material is based upon work supported by NOKIA and the NYU WIRELESS Industrial Affiliates Program, three National Science Foundation (NSF) Research Grants: 1320472, 1302336, and 1555332, and the GAANN Fellowship Program. This work is also supported by the National Instruments Lead User Program. The authors thank Y. Xing, J. Koka, R. Wang, and D. Yu for their help in conducting the measurements. G. R. MacCartney, Jr. (email: gmac@nyu.edu), H. Yan, S. Sun, and T. S. Rappaport, are with the NYU WIRELESS Research Center, NYU Tandon School of Engineering, New York University, Brooklyn, NY 11201.}
}

\maketitle
\begin{tikzpicture}[remember picture, overlay]
\node[font=\small] at ($(current page.north) + (0,-0.2in)$) {G. R. MacCartney, Jr., H. Yan, S. Sun, and T. S. Rappaport, ``A Flexible Wideband Millimeter-Wave Channel Sounder with Local Area and NLOS to LOS};
\node[font=\small] at ($(current page.north) + (0,-0.35in)$)  {Transition Measurements," in \textit{2017 IEEE International Conference on Communications (ICC)}, Paris, France, May 2017, pp. 1-7.};
\end{tikzpicture}

\begin{abstract}
This paper presents a millimeter-wave (mmWave) wideband sliding correlator channel sounder with flexibility to operate at various transmission rates. The channel sounder can transmit and receive up to 1 GHz of RF null-to-null bandwidth while measuring a 2 nanosecond multipath time resolution. The system architecture takes advantage of field-programmable gate arrays (FPGAs), high-speed digital-to-analog converters (DACs), and low phase noise Rubidium (Rb) references for synchronization. Using steerable narrowbeam antennas, the system can measure up to 185 dB of path loss. The channel sounder is used to measure the directional and omnidirectional received power as a receiver transitions from line-of-sight to non-line-of-sight conditions down an urban canyon. A 25 dB drop in omnidirectional received power was observed as the receiver transitioned from line-of-sight (LOS) conditions to deeply shadowed non-LOS (NLOS) conditions. The channel sounder was also used to study signal variation and spatial consistency for a local set of receiver locations arranged in a cluster spanning a 5 m x 10 m local area, where the omnidirectional received power in LOS and NLOS environments is found to be relatively stable with standard deviations of received power of 2.2 dB and 4.3 dB, respectively. This work shows that when implementing beamforming at the transmitter at mmWave, the omnidirectional received power over a local area has little fluctuation among receiver locations separated by a few to several meters. 
\end{abstract}

\iftoggle{conference}{}{
}

\section{Introduction}\label{sec:intro}
The heavily occupied frequencies below 6 GHz have motivated the wireless industry to explore the use of higher frequencies such as mmWaves, which offer much wider bandwidths and a tremendous opportunity for the development of consumer, vehicular, and industrial wireless applications~\cite{Pi11a,Boccardi14a,Rap14a}. The Federal Communications Commission (FCC) in the United States (U.S.) offered the world's first rules for using mmWave frequencies in July 2016, setting the stage for the U.S. and the world to innovate and create new wireless technologies~\cite{FCC16-89}. 

The \textit{3rd Generation Partnership Project (3GPP)}, viewed as the global standard, has released its first channel model for frequencies above 6 GHz in TR 38.900 (Release 14), and is planning to complete this study item by 2017. However, the first official release of fifth-generation (5G) specifications (Release 15) is expected in 2018~\cite{3GPP.38.900}. Channel models cannot be developed unless vast amounts of measurements are collected using channel sounders. Channel sounders enable propagation measurements that generate the knowledge and understanding to characterize the wireless channel~\cite{Newhall96a,Newhall96b}, and have the ability to measure antenna patterns~\cite{Newhall97a}. The current 3GPP (Release 14) channel model~\cite{3GPP.38.900} was developed from ray-tracing simulations and numerous mmWave propagation measurements in a variety of scenarios such as urban macrocell (UMa), urban microcell (UMi), and indoor hotspot (InH) areas~\cite{A5GCM15,Haneda16a,Haneda16b,Sun16b,Mac15b,Rap15b}. 

More advanced channel sounders with much wider bandwidth (1 GHz or more) capabilities are necessary for mmWave channel modeling compared to lower frequency and narrower bandwidth systems that were used for sub-6 GHz channel modeling studies like WINNER II and the 3GPP (Release 12) channel model study for 4G and LTE~\cite{WinnerII,3GPP.36.873}. State-of-the-art systems assist in developing new air-interfaces, networks, and carrier deployments for much higher frequencies, wider bandwidths, and future cellular systems~\cite{Thoma92a}. Therefore, we present a new wideband sliding correlator channel sounder~\cite{Rap02aChap45} in Section~\ref{sec:Msys} of this paper, that operates across 1 GHz of radio frequency (RF) bandwidth and at many mmWave frequencies. Typical state-of-the-art mmWave channel sounders can be found in~\cite{Papazian16a,Salous16a,Wen16a,Park16a,Ben-Dor11a}. 

The measurement system was used for two site-specific measurement campaigns to learn about propagation characteristics as a receiver (RX) moves from LOS to NLOS, and also to observe omnidirectional received power at multiple RX locations separated by a few to several meters (m) in a local area. As wireless networks evolve with increasing complexity, understanding the spatial consistency of channel parameters is necessary, and the measurements and analysis in Sections~\ref{sec:Mproc} and~\ref{sec:Mresults} provide a glimpse of early results at mmWave, with concluding remarks in Section~\ref{sec:conc}.

\section{Measurement System and Hardware}\label{sec:Msys}
The wideband mmWave channel sounder is based on the sliding correlator architecture developed by D. Cox~\cite{Cox72a,Rap02aChap45}. The transmitter (TX) baseband was built with National Instruments (NI) hardware and the LabVIEW and LabVIEW-FPGA development environments. A PXIe-1085 chassis on a TX cart contains mounted hardware modules such as the PXIe-8135 Host Controller that runs Windows 7, a PXIe-6674T for FPGA and DAC clock distribution, and a NI PXIe-7966R FPGA card with an Active Technologies (AT)-1120 FPGA Adapter Module (FAM) DAC mounted to it. A stand-alone FS725 Rb frequency reference module with a 10 MHz output is connected to the chassis front-panel and is distributed to each module along the backplane of the chassis.

\subsection{Baseband and Probing Signal}
A pseudorandom noise (PN) sequence of 2047 chips in length is generated via LabVIEW-FPGA on the NI PXIe-7966R card and output via the AT-1120 FPGA FAM DAC. The FPGA and DAC are programmed to generate PN sequence chip rates up to 500 megachips-per-second (Mcps), with an 11-bit linear feedback shift register (LFSR) state machine that uses a leap-forward (LF) LFSR architecture through digital logic~\cite{Chu99a}. The LFSR begins with a value of ``1" in all states with the 11\textsuperscript{th} and 9\textsuperscript{th} taps exclusive-or'd and fed back to the first tap. A leap forward LFSR is used for generating high rate PN codes since the FPGA and FAM DAC are clocked at a fixed rate of 125 MHz, where 16 time-interleaved channels are accessible in each single-cycle timed loop (SCTL), which allows for a (125 MHz x 16) 2 Giga-Samples-per-second (GS/s) sample rate. This architecture is a common aspect of serializers and deserializers (SerDes). For instance, to generate a 500 Mcps PN sequence, 4 samples per chip and 4 total chips are required by the DAC in each SCTL, with a new set of 4 chips required in each successive loop. The LFSR LF state-machine ``leaps-forward" several states between successive loops so as to have access to enough chips to generate the appropriate transmission signal. The LFSR LF state machine is configurable for various PN codes and lengths. 

\begin{figure}[b!]
	\centering
	\includegraphics[width=0.37\textwidth]{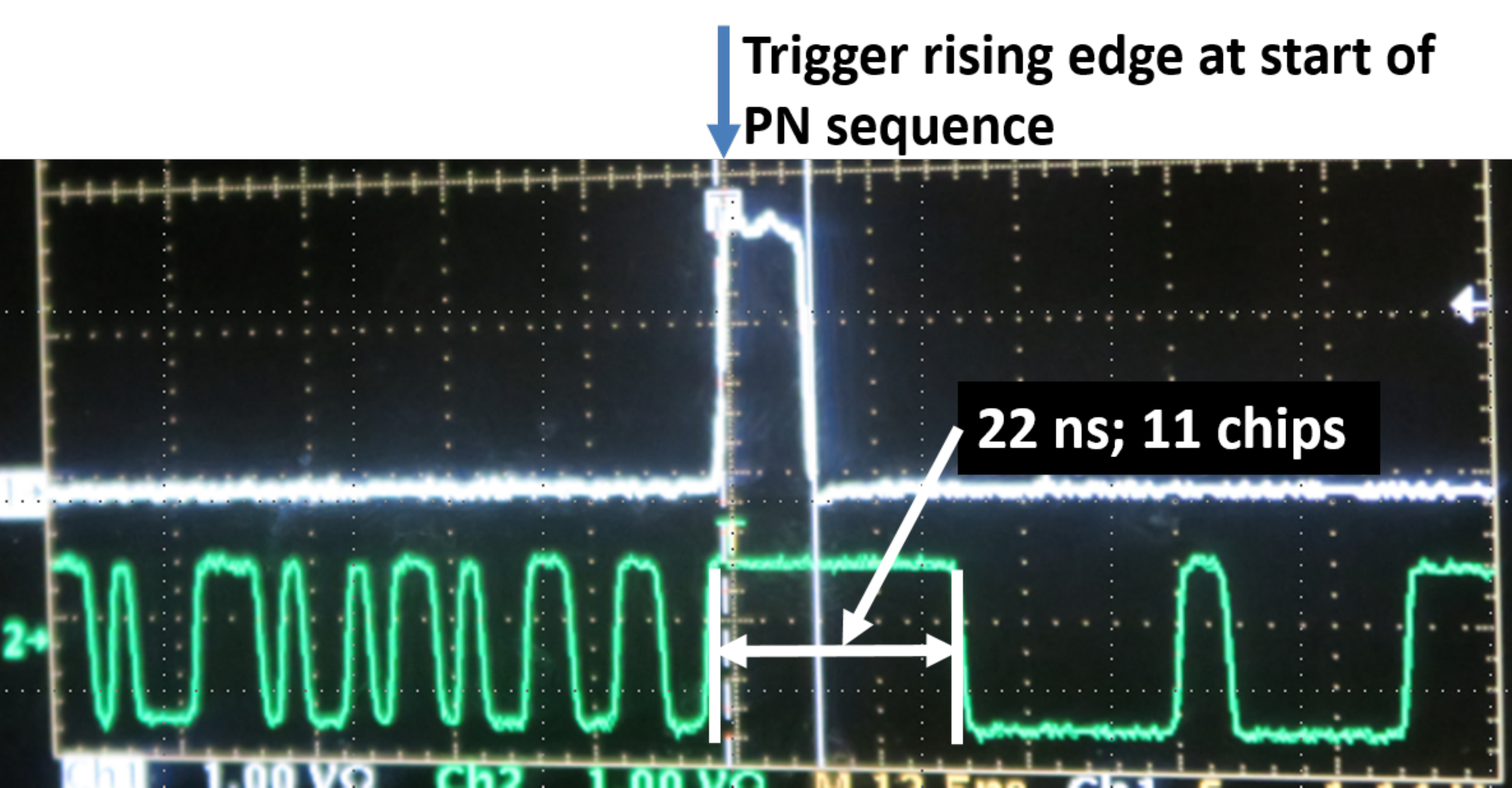}
	\caption{500 Mcps PN sequence (code length $2^{11}-1=2047$, starting with 11 ones) with digital trigger (8 nanosecond width) at start of sequence.}
	\label{fig:PN_trig}
\end{figure}

An advantage of the DAC and FPGA code generation is the ability to output a trigger at the start time of the PN sequence for synchronization between the TX and RX. Fig.~\ref{fig:PN_trig} shows an image of the start of the PN sequence (11 ``1's") and the trigger aligned with the start of the code (tiny cable delay), that can be flexibly shifted around in 8 ns increments (FPGA clock period of $\frac{1}{125\;\MHz} = 8$ ns) to account for cable and system delays.
\begin{figure}[b!]
	\centering
	\includegraphics[width=0.22\textwidth]{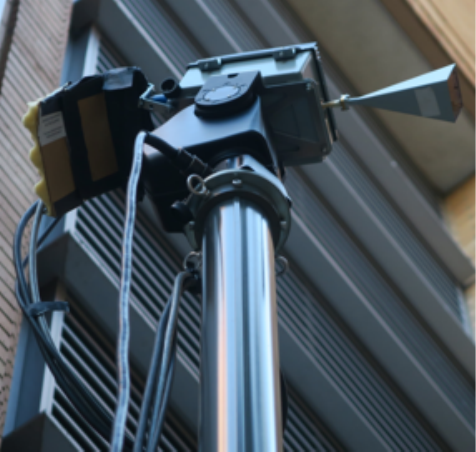}
	\caption{RF upconverter with horn antenna, mounted on a gimbal, and atop a 4 meter high mast.}
	\label{fig:TX_horn}
\end{figure}

\subsection{TX Superheterodyne}
\begin{figure*}[t!]
	\centering
	\includegraphics[width=0.85\textwidth]{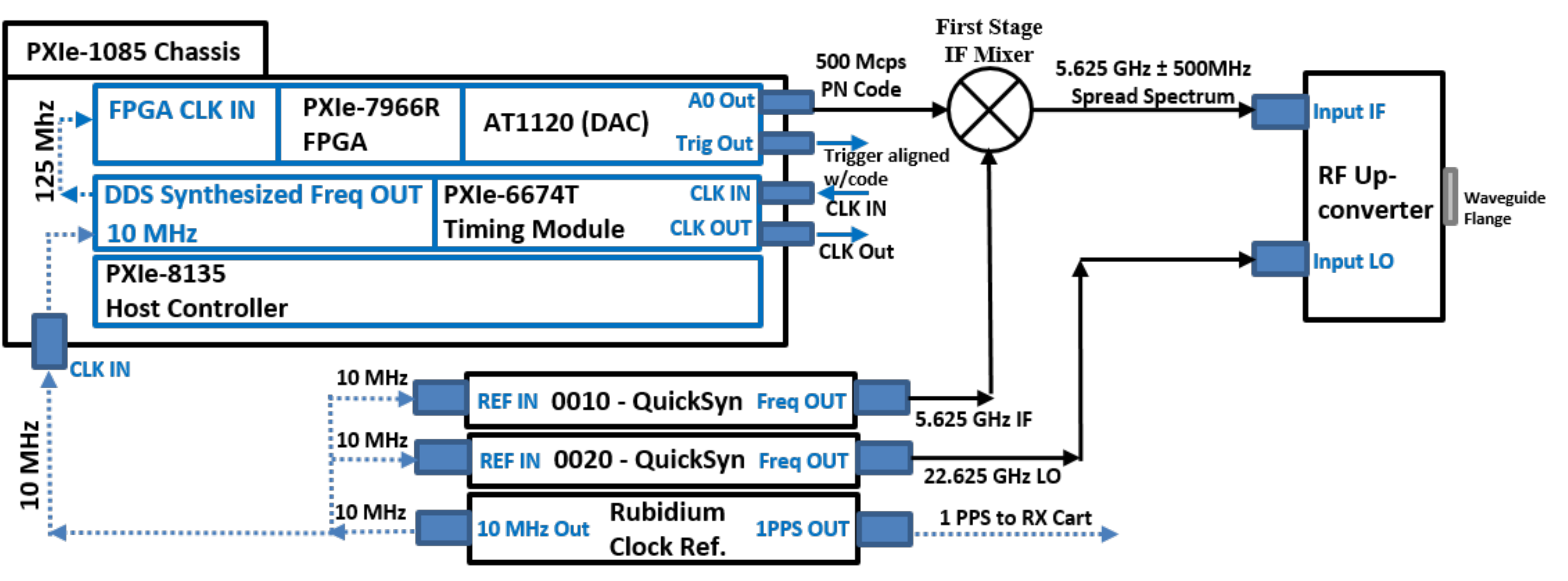}
	\caption{Channel sounder transmitter system block diagram~\cite{Rap13a,Nie13a,Mac14a,Mac17a}.}
	\label{fig:TX_BD}
\end{figure*}

The baseband signal is modulated with an intermediate-frequency (IF) of 5.625 GHz that is generated via an NI FSW-0010 QuickSyn frequency synthesizer (FS) that has its 10 MHz ``Ref in" supplied from the FS725 Rb clock. The 1 GHz wide null-to-null bandwidth signal at IF then enters a custom RF upconverter (additional information provided in~\cite{Rap13a,Nie13a,Mac14a}). The local oscillator (LO) supplied to the upconverter box is 22.625 GHz and is generated via a customized FSW-0020 QuickSyn FS (with frequency doubler) with its 10 MHz ``Ref in'' supplied from the FS725. The RF upconverter triples the LO input to 67.875 GHz for upconversion from IF to an RF center frequency of 73.5 GHz. The TX schematic is given in Fig.~\ref{fig:TX_BD}~\cite{Nie13a,Mac14a}.
\subsection{TX Antenna Control}
The 73 GHz upconverter has a waveguide flange output that allows for horn antennas to be mounted as shown in Fig.~\ref{fig:TX_horn}. To maneuver the azimuth and elevation angles of departure (AODs) and angles of arrival (AOAs) with fine precision, the RF upconverter is mounted to a FLIR Pan-Tilt D100 gimbal that allows for sub-degree control through a custom-programmed LabVIEW graphical user interface (GUI) and with a Sony Play Station 3 (PS3) game controller (see Fig.~\ref{fig:PS3})~\cite{Mac14a}. Also shown in Fig.~\ref{fig:TX_horn} is the upconverter and gimbal attached to a pneumatic mast that can be elevated to a height of 4 meters (m) to simulate lamppost base station heights in typical UMi scenarios.

\subsection{RX Superheterodyne}
\begin{figure*}
	\centering
	\includegraphics[width=0.9\textwidth]{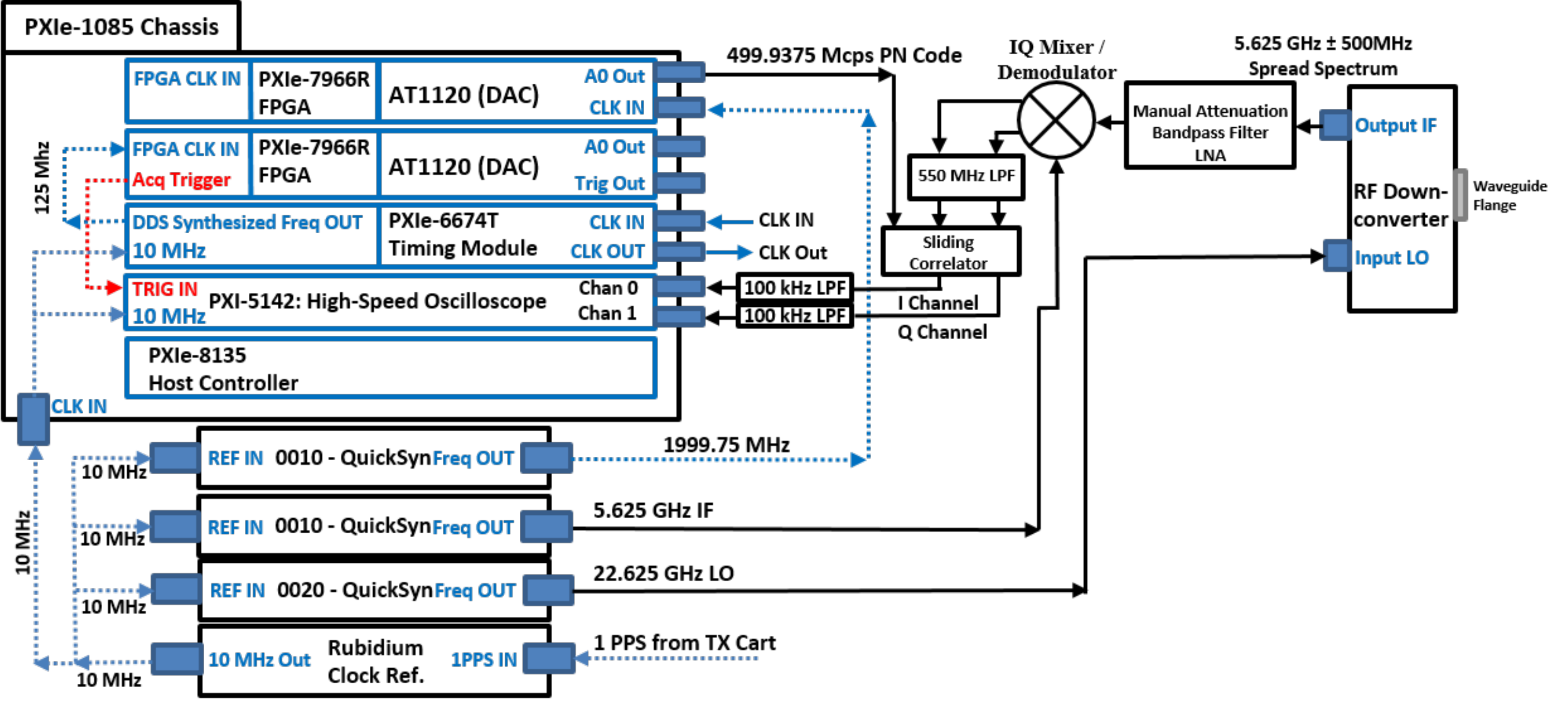}
	\caption{Channel sounder receiver sliding correlator block diagram~\cite{Rap13a,Nie13a,Mac14a,Mac17a}.}
	\label{fig:RX_BD}
\end{figure*}

A horn antenna captures the incoming RF signal at 73.5 GHz at the RX, which then enters a waveguide flange input on the downcoverter~\cite{Mac14a,Nie13a}. The RF downconverter is fed an LO input of 22.625 GHz via an FSW-0020 QuickSyn FS (with frequency doubler) that is synced to a 10 MHz clock from a second FS725 module, at the RX cart. The LO is frequency tripled to 67.875 GHz in order to downconvert the RF signal to an IF centered at 5.625 GHz. The IF signal then enters a manual step attenuator (0-110 dB) for gain control, followed by an IF bandpass filter after-which the signal is then amplified by a low-noise-amplifier (LNA) and then demodulated into its in-phase ($I$) and quadrature-phase ($Q$) baseband components. The $I$ and $Q$ baseband signals are then low pass filtered slightly above the TX chip rate to remove aliasing. The $IQ$ demodulation mixer is fed an LO of 5.625 GHz via an FSW-0010 QuickSyn FS that has its ``Ref In" locked to the FS725 10 MHz clock. Fig.~\ref{fig:RX_BD} shows the RX system and baseband processing for the sliding correlator that is described next. 
\subsection{RX Sliding Correlator}
The sliding correlator method is configurable for various chip rates, but the following description and diagram is for a 500 Mcps rate. A PXIe-1085 chassis contains mounted NI hardware modules at the RX cart and is fed a 10 MHz reference clock from the RX FS725, which is then distributed along the backplane of the chassis. An identical PN generator is used at the RX to generate a slightly slower RX chip rate of 499.9375 Mcps, by slightly modifying the FPGA clock rate and DAC module clock. An FSW-0010 Quicksyn FS (synced to the RX FS725) is used to generate a 1999.75 MHz clock which is divided by 4 to result in a chip rate of 499.9375 Mcps. 

The sliding correlation process of mixing the received wideband signal and a local copy at the RX results in the $I$ and $Q$ channel impulse response (CIR) signals, which are bandwidth compressed to 62.5 kHz: $500\;\MHz - 499.9375\;\MHz=62.5$ kHz. The bandwidth compressed $I$ and $Q$ signals are subsequently low pass filtered and sampled simultaneously via a PXI-5142 two-channel high-speed oscilloscope. The PXI-5142 receives its digital trigger via the chassis backplane from a PXIe-7966R FPGA, that is periodic with the time-dilated CIR. The trigger time is adjustable and allows the operator to manually shift the CIR to have a desired time delay. The PXI-5142 receives its 10 MHz reference from the chassis backplane.

The period of the time-dilated CIR is:
\begin{equation}\small
2047\times\frac{1}{500\,\MHz-499.9375\,\MHz} = 2047\times\frac{1}{62.5\,\kHz} = 32.752\;\mathrm{ms}
\end{equation} 
and corresponds to a time-dilation factor (slide factor~\cite{Rap02aChap45}) of $8\,000$, which results in a processing gain of 39 dB~\cite{Rap02aChap45}. During the recording process, 20 consecutive periods of the $I$ and $Q$ voltage channel CIRs are individually acquired, squared, and then summed together ($I^2+Q^2$) to form 20 power delay profiles (PDPs). The 20 PDPs are then non-coherently averaged together (in linear) in order to improve signal-to-noise ratio (SNR) by approximately $10\log_{10}(\sqrt{20})=6.5$ dB~\cite{Skolnik01a}. It is worth noting that the 73 GHz sliding correlator system can measure a maximum path loss of approximately 185 dB when operating with a maximum TX RF output power of 14.6 dBm, with 27 dBi horn antennas at the TX and RX, and with a 5 dB SNR noise floor threshold~\cite{Rap15b}. 

\subsection{RX Antenna Control, Recording Software, and Synchronization}\label{sec:RXcontrol}
\begin{figure}
	\centering
	\includegraphics[width=0.46\textwidth]{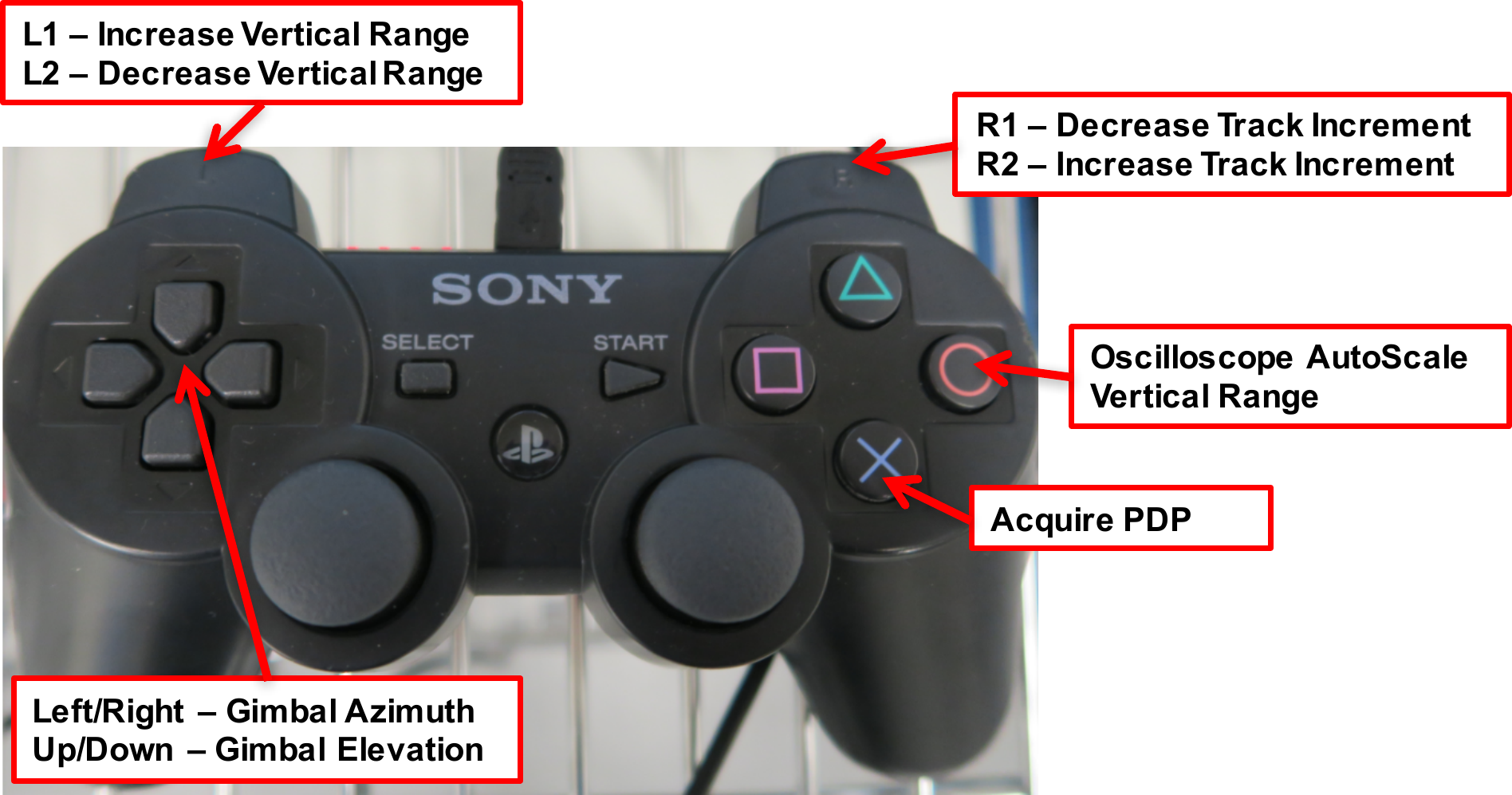}
	\caption{Channel sounder acquisition and antenna positioning controller.}
	\label{fig:PS3}
\end{figure}

Similar to the TX, the RX downconverter and antenna are attached to a rotatable gimbal that is secured to either a 14-inch length linear track for small-scale measurements, a tripod, or pneumatic mast. The gimbal at the RX is controlled via a PS3 game controller shown in Fig.~\ref{fig:PS3}, in order to maneuver the RX antenna azimuth and elevation angles. The PS3 game controller may also be used for acquiring PDPs, translating the linear track, and adjusting the oscilloscope vertical scale. Additionally, the TX antenna can be controlled from the RX via an ad hoc WiFi link in LabVIEW. 

Acquisition software was written in LabVIEW to view live PDPs, to save and log all recorded PDPs, for power calibration, and for timing calibration. The LabVIEW GUI allows for flexible control and the ability to ``kick-off" automated track and azimuth sweep measurements that record PDPs at discrete angles or positions, with real-time feedback of the channel. The FS725 Rb reference at the RX can receive a 1 PPS input signal from the TX FS725 output 1PPS for frequency reference synchronization. This step ensures that all frequency references in the system are synchronized well and do not experience significant drift. After an hour of training, the two FS725's can be disconnected and their 10 MHz outputs will slowly begin to drift apart. Single acquisitions take less than one second to record and are time coherent. Acquisitions at different angle orientations may experience drift between captures. The drift can be removed with post-processing techniques that allow for coherent excess delay measurements across all antenna angle orientations for a TX-RX location combination. This capability allows for constructing a complete PDP profile from all spatial angles at the TX and RX. Additional details regarding synchronization and drift can be found in~\cite{Mac17a}. The channel sounder system described here is easily interchanged with RF heads for various mmWave and sub-6 GHz frequency bands. The system specifications and measurement details are now described in Section~\ref{sec:Mproc}.

\section{Measurement Specifications and Procedure}\label{sec:Mproc}
The 73 GHz wideband sliding correlator channel sounder described in Section~\ref{sec:Msys} was used to conduct large-scale measurements in downtown Brooklyn, NY, at the MetroTech Commons courtyard near 2 \& 3 MetroTech Center. The maximum RF TX output power was 14.6 dBm with the use of a 27 dBi steerable horn antenna, with 7$^\circ$ azimuth and elevation half-power beamwidth (HPBW). The resulting effective isotropic radiated power (EIRP) was 41.6 dBm. Additional system specifications for the measurements are given in Table~\ref{tbl:MeasSys}. 
\begin{table}
	\centering
	\caption{Wideband sliding correlator channel sounder system specifications for the 73 GHz measurements.}
	\label{tbl:MeasSys}
	\begin{center}
		\scalebox{0.82}{
			\begin{tabu}{|c|[1.6pt]c|}
				\hline 
				Description &	Specification \\ \specialrule{1.5pt}{0pt}{0pt}
				Baseband Sequence &	PRBS (11th order: 2$^{11}$-1 =  Length 2047)	\\ \hline
				TX / RX Chip Rate &	500 Mcps / 499.9375 GHz	\\ \hline
				Slide Factor $\gamma$ &	$8\,000$	\\ \hline
				RF Null-to-Null Bandwidth &	1 GHz	\\ \hline
				Multipath Time Resolution &	2 ns	\\ \hline
				PDP Threshold &	20 dB down from max peak / +5 dB SNR	\\ \hline
				TX/RX Intermediate Frequency &	5.625 GHz	\\ \hline
				TX/RX LO &	67.875 GHz (22.625 GHz $\times 3$)	\\ \hline
				Synchronization &	TX/RX trained 10 MHz via 1PPS reference	\\ \hline
				Carrier Frequency &	73.5 GHz	\\ \hline
				Max TX RF Power &	14.6 dBm	\\ \hline
				TX/RX Antenna Gain &	27 dBi / 20 dBi	\\ \hline
				TX Azimuth and Elevation HPBW &	7$^\circ$	\\ \hline
				RX Azimuth and Elevation HPBW &	15$^\circ$	\\ \hline
				TX/RX Antenna Polarization &	V-V	\\ \hline
				EIRP &	41.6 dBm	\\ \hline
				TX/RX Heights &	4 m / 1.5 m	\\ \hline
		\end{tabu}}
	\end{center}
\end{table}
\subsection{Measurement Procedure and Scenarios}
A route based measurement scenario (case one) was tested with a fixed TX location and 16 RX locations, where the RX locations were adjacently separated by 5 meters. The path of adjacent RX locations was used to simulate a mobile receiver moving along a trajectory from LOS conditions (five locations) to NLOS conditions (11 locations). A second measurement scenario (case two) used a fixed TX and a set of closely separated RX locations (a few to several meters) arranged in a cluster, where one cluster consisted of five RX locations in LOS and the second cluster consisted of five RX locations in NLOS.

\subsection{Measurement Procedure}
The procedures for both measurement scenarios were similar, where the TX and RX antennas were set to 4.0 m and 1.5 m above ground level (AGL), respectively. For each TX-RX combination tested, five consecutive and identical azimuth sweeps ($\sim$3 minutes per sweep and $\sim$2 minutes between sweeps) were conducted at the RX in HPBW step increments (15$^\circ$) where a PDP was recorded at each RX antenna azimuth pointing angle. This procedure resulted in at most 120 PDPs $\left(\frac{360^\circ}{15^\circ}\times 5=120\right)$ per TX-RX combination (some angles did not have detectable signal above the noise). The TX antenna azimuth and elevation pointing angles were fixed for all RX locations for the route measurements, whereas the TX antenna azimuth and elevation pointing angles were fixed separately for each of the two cluster based measurements. For each RX location measured for both scenarios, the best RX antenna pointing angle in the azimuth was selected as the starting point for each RX azimuth sweep, whereas the RX antenna elevation angle remained fixed for the entire set of route measurements and then for each of the two cluster measurements.  

\subsection{Measurement Scenarios}
\subsubsection{Case One -- Route Measurements}
\begin{figure}
	\begin{center}
		\includegraphics[width = 0.44\textwidth]{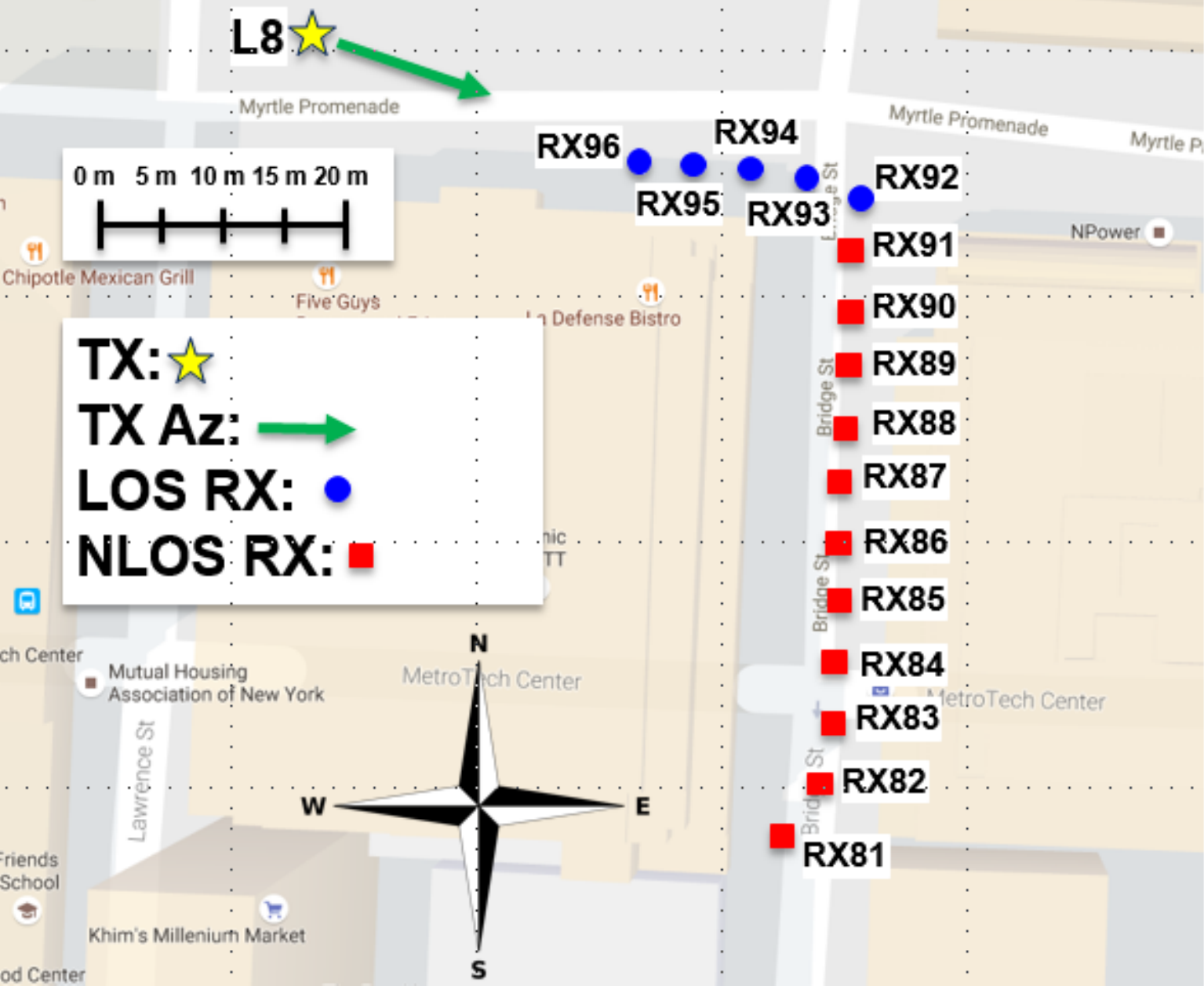}
		\caption{2D map of TX and RX locations for the route (case one) LOS to NLOS transition measurements. The yellow star is the TX location, blue dots represent LOS RX locations, and red squares indicate NLOS RX locations.}\label{fig:Map_Large_Scale_C1}
	\end{center}
\end{figure}

For the route measurements, 16 RX locations were measured for a fixed TX location (L8) with the RX locations set in 5 m adjacent increments that simulated a route in the shape of an ``L" around a building corner from a LOS to NLOS region as shown in Fig.~\ref{fig:Map_Large_Scale_C1}. 
The Euclidean (straight line on a map) T-R separation distances varied from 29.6 m to 49.1 m for the LOS locations (RX96 to RX92) and from 50.8 m to 81.6 m for the NLOS locations (RX91 to RX81). At L8, the TX antenna azimuth/elevation departure angles were fixed to 100$^\circ$/0$^\circ$ for all 16 RX locations measured. The RX antenna elevation was fixed at 0$^\circ$ for all 16 RX locations. Therefore, it is important to note that the LOS measurements by nature do not include a perfect boresight-to-boresight angle alignment between the TX and RX antennas. The general layout of measurements consisted of the RX location starting at RX96, which is in clear LOS view of the TX, aside from minor foliage and lamppost obstructions nearby. The TX antenna was pointed towards the opening of the urban canyon for all RX locations (see Fig.~\ref{fig:Map_Large_Scale_C1}). The last NLOS location at RX81 is approximately 54 m down the urban canyon (street width of $\sim$18 m).

\subsubsection{Case Two -- Cluster Measurements}
\begin{figure}
	\begin{center}
		\includegraphics[width = 0.46\textwidth]{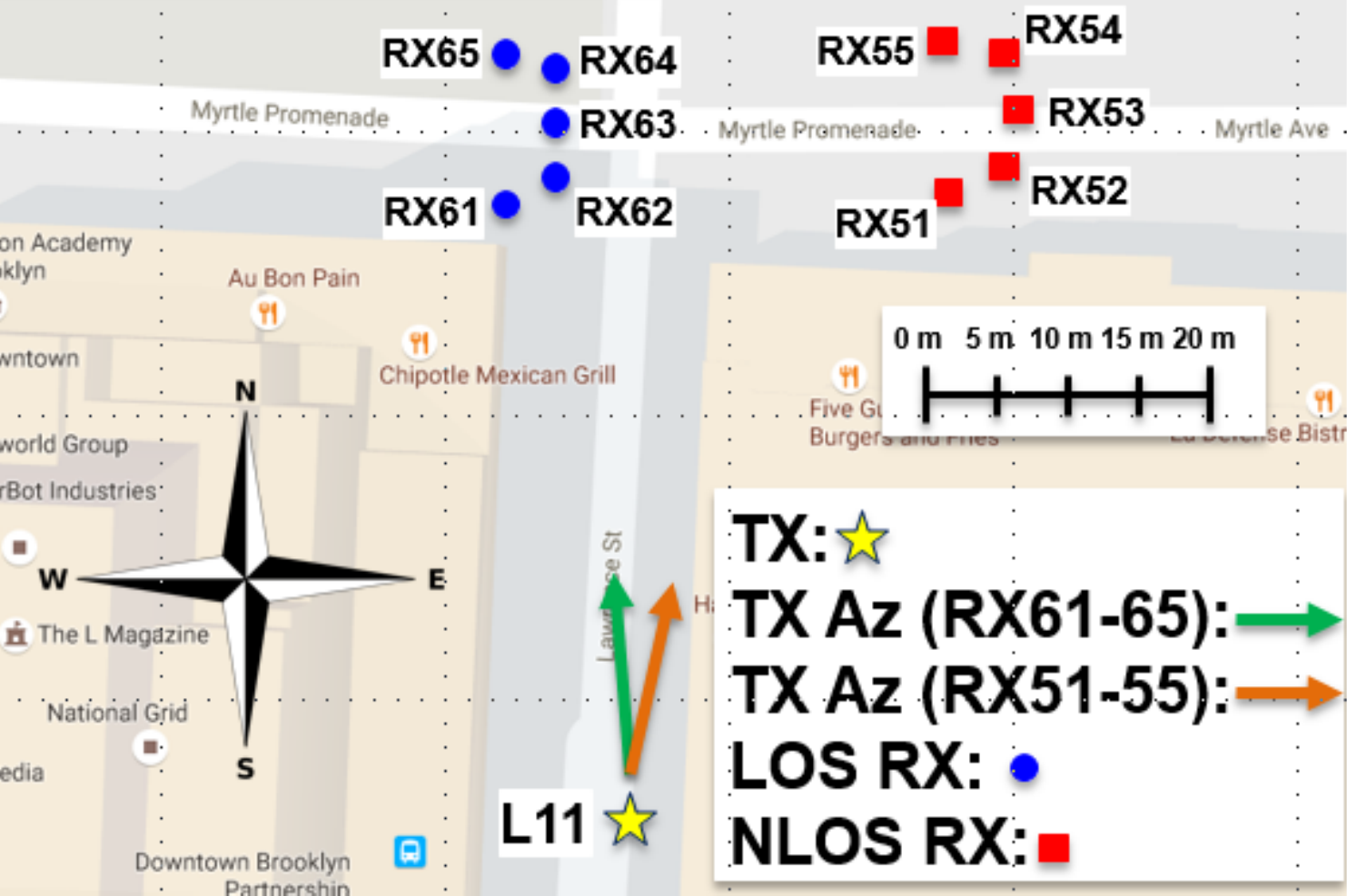}
		\caption{2D map of TX and RX locations for the cluster (case two) measurements with LOS and NLOS RX clusters. The yellow star is the TX location, blue dots represent LOS RX locations, and red squares indicate NLOS RX locations.}\label{fig:Map_Large_Scale_C2}
	\end{center}
\end{figure}

For the cluster measurements, 10 RX locations were measured for a fixed TX location (L11), with two sets of RX clusters, one in LOS (RX61 to RX65) and the other in NLOS (RX51 to RX55). For each cluster of RX's, the distance between each adjacent RX location was 5 m, whereas the path of adjacent RX locations took the shape of an arc, as displayed in Fig.~\ref{fig:Map_Large_Scale_C2}. The LOS cluster Euclidean T-R separation distances varied between 57.8 m and 70.6 m with a fixed TX antenna azimuth/elevation departure angle of 350$^\circ$/-2$^\circ$, where each RX had a fixed antenna elevation angle of $+3^\circ$ above the horizon. The Euclidean T-R separation distances for the NLOS cluster are between 61.7 m and 73.7 m with a fixed TX antenna azimuth/elevation departure angle of 5$^\circ$/-2$^\circ$. The RX antenna elevation was fixed to $+3^\circ$ for each of the five NLOS locations. The LOS cluster of RX's was located near the opening of an urban canyon near some light foliage, while the TX location was $\sim$57 m down an urban canyon (street width of $\sim$18 m). The NLOS cluster of RX locations was around the corner of the urban canyon opening, with light to moderate foliage and lampposts nearby.

\begin{figure}
	\begin{center}
		\includegraphics [width = 0.45\textwidth]{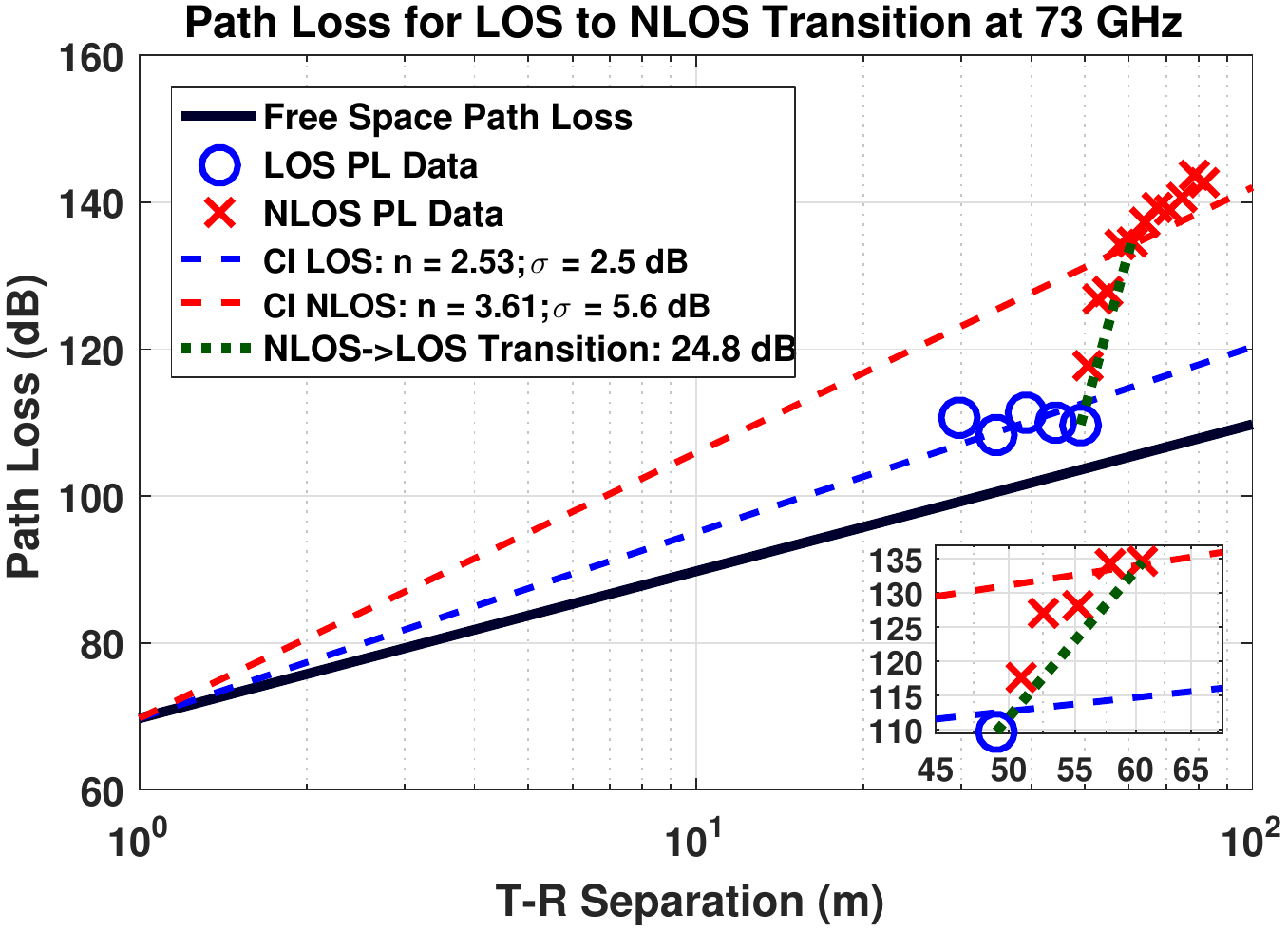}
		\caption{Omnidirectional path loss and CI models with a 1 m reference distance for the large-scale route measurements with an RX transitioning from LOS to NLOS down an urban canyon.}\label{fig:routePL}
	\end{center}
\end{figure} 

\begin{figure}
	\centering
	\begin{subfigure}[b]{0.45\textwidth}
		\centering
		\includegraphics[width=\textwidth]{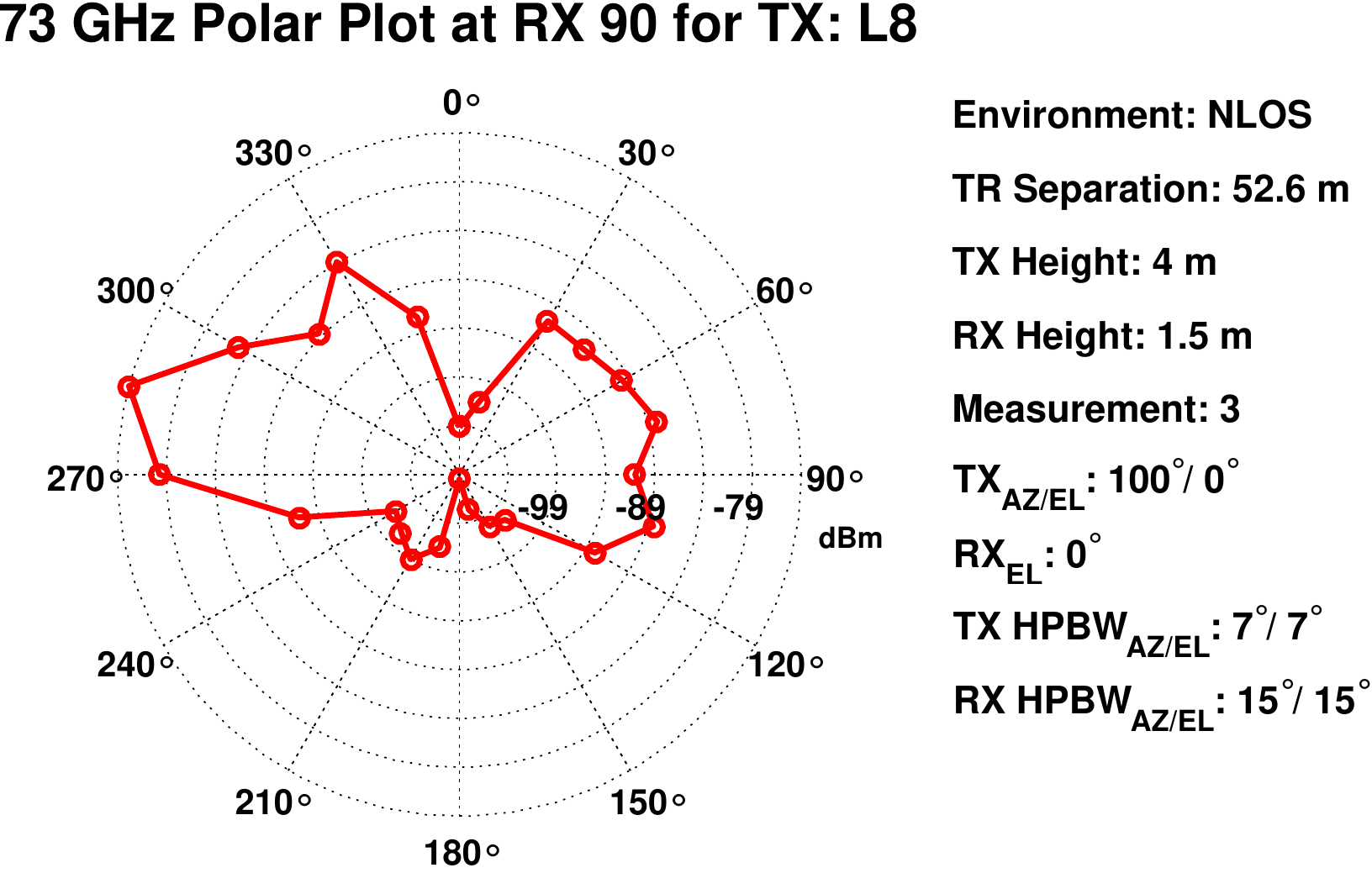}
		\caption[Network2]%
		{{\small RX90: NLOS}}   
		\label{fig:RX90}
	\end{subfigure}
	\hfill
	\begin{subfigure}[b]{0.45\textwidth}  
		\centering 
		\includegraphics[width=\textwidth]{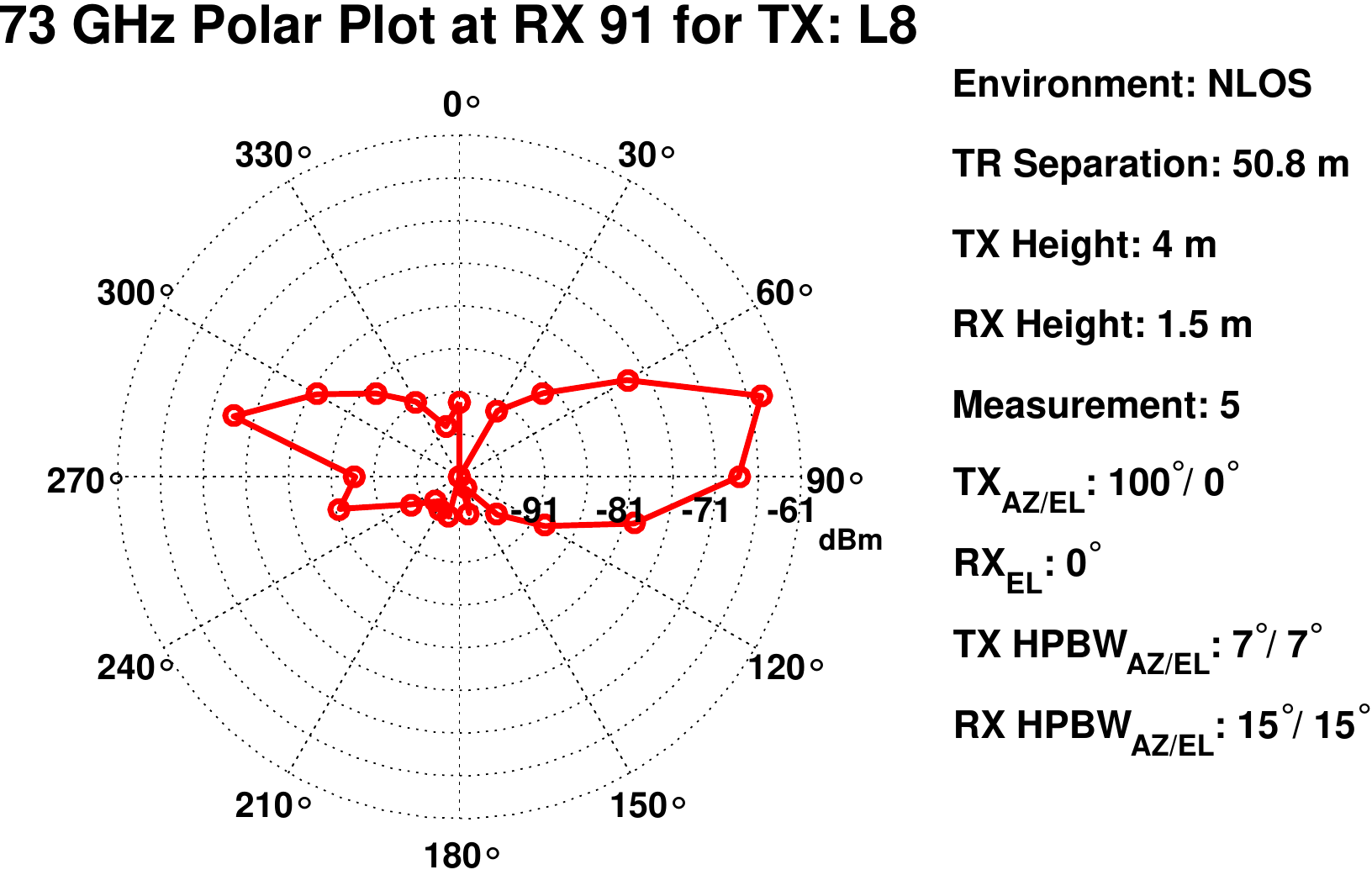}
		\caption[]%
		{{\small RX91: NLOS}}    
		\label{fig:RX91}
	\end{subfigure}
	\hfill
	\begin{subfigure}[b]{0.45\textwidth}   
		\centering 
		\includegraphics[width=\textwidth]{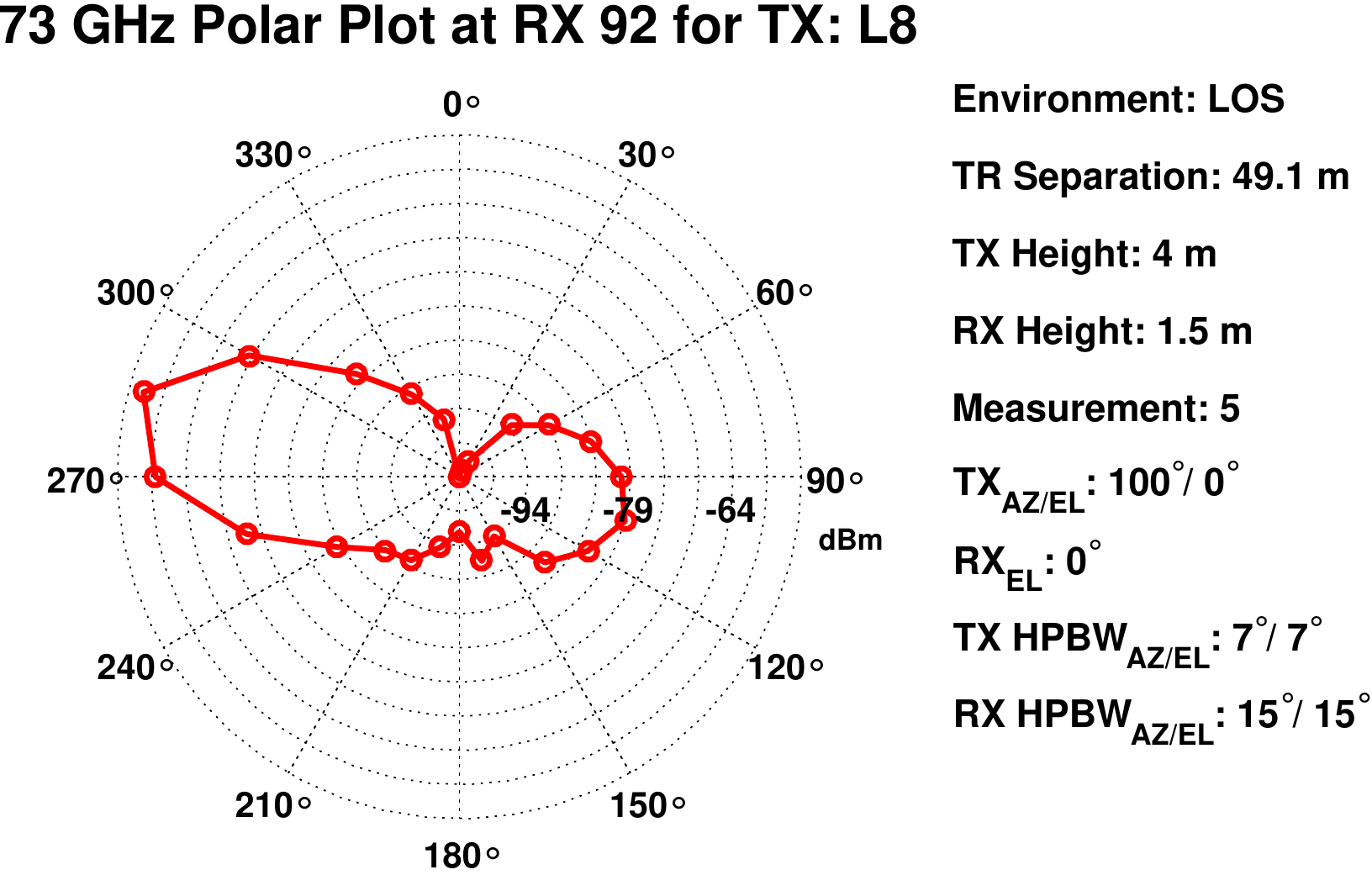}
		\caption[]%
		{{\small RX92: LOS}}    
		\label{fig:RX92}
	\end{subfigure}
	\hfill
	\begin{subfigure}[b]{0.45\textwidth}   
		\centering 
		\includegraphics[width=\textwidth]{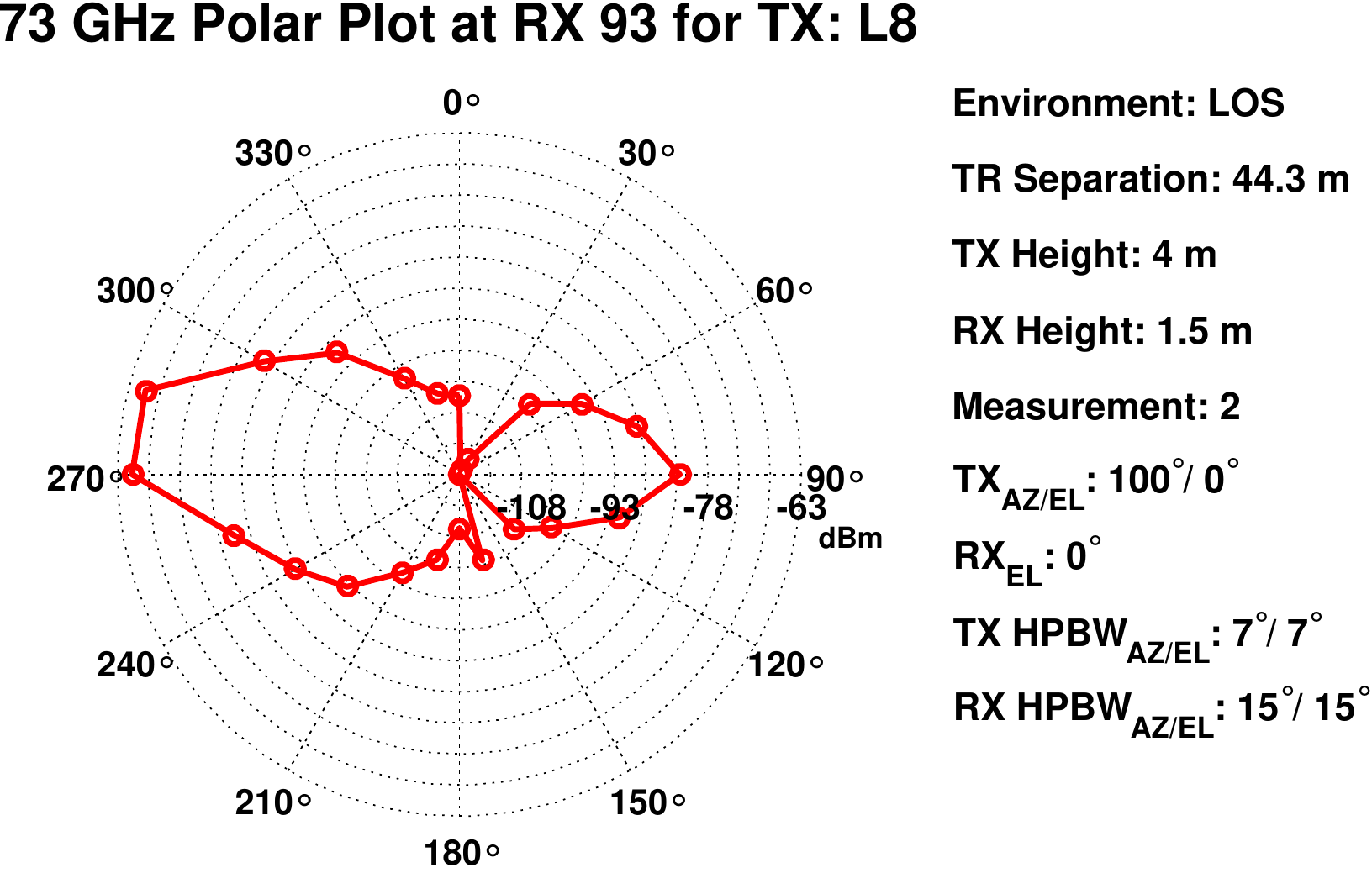}
		\caption[]%
		{{\small RX93: NLOS}}    
		\label{fig:RX93}
	\end{subfigure}
	\caption[ The average and standard deviation of critical parameters ]
	{\small Route-based (case one) polar plots of RX azimuth spectra for two NLOS locations (RX90 and RX91) and two LOS locations (RX92 and RX93) that show the evolution of the AOA energy around a corner.} 
	\label{fig:case1Polar}
\end{figure}

\section{Measurement Results and Analysis}\label{sec:Mresults}
The route measurements mimicked that of a mobile receiver moving around a corner and down an urban canyon from LOS conditions to NLOS conditions, to understand the evolution of the channel as an RX makes such a transition. Fig.~\ref{fig:routePL} displays the omnidirectional path loss for each of the RX locations (RX81 to RX96) where the received powers from the individual directional measurements were summed up to determine the entire omnidirectional received power (out of the 5 sweeps, the maximum power at each angle was used), following the procedure in~\cite{Mac14b,Sun15a}.

The transition from LOS to NLOS in Fig.~\ref{fig:routePL} is quite abrupt, where path loss increases initially by $\sim$8 dB from RX92 to RX91. The received power at RX90 is 9 dB less than at RX91, and then subsequently drops 1 dB more from RX90 to RX89, 6 dB more from RX89 to RX88, and then a 1 dB drop from RX88 to RX87, approximately half-way down the urban canyon. This corresponds to an overall 25 dB increase in path loss when transitioning from a LOS area to a deeply shadowed NLOS location. The path distance from RX92 to RX88 is 20 meters, indicating that the received signal strength dropped by approximately 1.25 dB/m. This fading rate will be important for developing handoff algorithms at mmWave, where vehicle speeds of 35 m/s will experience a signal fading rate of 44 dB/s, whereas a mobile at a walking speed of 1 m/s would experience a fading rate of 1.25 dB/s. The close-in free space reference distance (CI) path loss model with a 1 m reference distance was determined for the LOS and NLOS locations, separately~\cite{Sun16b}. The CI model path loss exponent (PLE) of 2.53 in LOS was found to be higher than free space (n=2), but can be attributed to the TX and RX antennas only roughly aligned on boresight, since the test was conducted for a fixed TX antenna azimuth/elevation departure angle. Furthermore, the relatively constant path loss observed from RX92 to RX96 is likely due to the mismatch in elevation alignment between the TX and RX antennas. The NLOS CI PLE of 3.61 down the urban canyon is slightly larger than previously reported measurements (NLOS $n=3.4$~\cite{Mac14b}), but is comparable since the measurement procedure was highly specific and did not consider various angles of departure. 

Azimuth power spectra are important for understanding how the arriving signal spatially transforms as an RX moves from LOS to NLOS. Fig.~\ref{fig:case1Polar} displays four polar plots, one each from RX90, RX91, RX92, and RX93, with the TX pointing approximately in the 285$^\circ$ direction at the LOS RX locations (see Fig.~\ref{fig:Map_Large_Scale_C1}). In Fig.~\ref{fig:RX90} it is noticeable that the signal arrives at the RX at two main lobes (centered at 285$^\circ$ and 75$^\circ$) with energy diffracting around the corner of the building and through a small archway at the corner of the building from the 285$^\circ$  AOA direction. The second lobe at RX90 in Fig.~\ref{fig:RX90} is centered near 75$^\circ$ and is from reflections off of the building to the east of the receiver, demonstrating the surprisingly reflective nature of the channel at the 73 GHz mmWave band~\cite{Samimi13a,Mac14a} in NLOS.

Fig.~\ref{fig:RX91} is the last NLOS RX (RX91) before LOS and it appears that a majority of the received power comes from the backside lobe facing the 75$^\circ$ AOA, likely from reflected energy off of the building to the east of RX91. The 285$^\circ$ RX91 AOA lobe is weaker than at RX90 and can be attributed to a stone pillar obstructing rays from the building to the west. As expected, the two LOS locations (RX92 and RX93) have main directions of arrival at 285$^\circ$ from where the TX antenna is pointing. Similar to RX90 and RX91, reflected energy off of the building to the east of RX92 and RX93 contributes to considerable power in a secondary lobe. Table~\ref{tbl:clusterSTD} displays the standard deviation of omnidirectional received power for the route measurement LOS and NLOS RX locations. The standard deviation of received power is 1.2 dB for the local set of LOS measurements. The standard deviation of omnidirectional received power is much larger in NLOS (7.9 dB), possibly due to scattered, weakly diffracted, and reflected energy received at the NLOS RX locations after the initial transition, compared to the much lower received power farther down the urban canyon.

\begin{table}
\centering
\caption{Large-scale route and cluster standard deviation of received power for local measurements}
\label{tbl:clusterSTD}
\begin{center}
	\scalebox{0.84}{
		\begin{tabu}{|c|[1.6pt]c|}
			\hline 
			\textbf{Measurement Set} & \textbf{Omnidirectional Received Power \bm{$\sigma$} [dB]} \\ \specialrule{1.5pt}{0pt}{0pt}
			Case 1 (\textit{Route}) - LOS: RX92 to RX96 &	1.2	\\ \hline
			Case 1 (\textit{Route}) - NLOS: RX81 to RX91 &	7.9	\\ \hline
			Case 2 (\textit{Cluster}) - LOS: RX61 to RX65 &	2.2	\\ \hline
			Case 2 (\textit{Cluster}) - NLOS: RX51 to RX55 &	4.3	\\ \hline
	\end{tabu}}
\end{center}
\end{table}

The case two \textit{cluster} measurements for the L11 TX were designed to understand how received power changes in local areas (larger than small-scale) on the order of many hundreds to thousands of wavelengths at mmWave. Under LOS conditions, the cluster of five RX locations over a 5 m x 10 m grid resulted in an omnidirectional received power standard deviation of 2.2 dB, relatively small, indicating consistent average received energy over the local set of RX locations in LOS. The NLOS cluster resulted in an omnidirectional received power standard deviation of 4.3 dB, and while it is approximately 2 dB larger than the LOS standard deviation, it is not as high as previously observed multi-frequency shadow fading in NLOS at mmWave, which is reported as 8.0 dB for UMi in~\cite{Sun16b}. The small fluctuation in omnidirectional received power over the local area NLOS cluster implies that received power does not change much over RX locations separated by a few to several meters in NLOS at mmWave, when using a fixed TX antenna beam. These observations of low fading variations of omnidirectional received power in LOS and NLOS at 73 GHz over local areas, are an early first step in empirical studies of spatial consistency at 73 GHz mmWave.

\section{Conclusion}\label{sec:conc}
This paper introduced a new mmWave channel sounder based on FPGAs, high-speed DACs, Rubidium standard frequency references, and acquisition software for conducting mmWave propagation measurements over various bandwidths and carrier frequencies. The wideband sliding correlator channel sounder is capable of transmitting codes with up to 1 GHz of RF null-to-null bandwidth and has a minimum multipath time resolution of 2 ns.  Measurements for a cluster of RX's for a fixed TX narrowbeam antenna pointing direction showed that omnidirectional received power remained relatively stable over five LOS locations (2.2 dB standard deviation) with 5 m distances between adjacent locations. NLOS cluster measurements resulted in a larger standard deviation of 4.3 dB, which is still relatively consistent for a fixed beam TX pointing direction in regards to SNR and signal variation over a 5 m x 10 m local area. A route-based measurement scenario for an RX with a transition from LOS to NLOS showed an abrupt 25 dB drop in omnidirectional received power over 20 meters. The measurements predict that a vehicle moving at 35 m/s would experience a signal fading rate of 44 dB/s for a LOS to NLOS transition around a building corner. These transition scenarios and fading rates will be important to consider when designing 5G handoff algorithms. This paper and work in~\cite{Sun17a} provide insights into the spatial consistency of received signal strength for mmWave communications.

\bibliography{../../../../../../../LaTeX/MacCartney_Bibv5}
\bibliographystyle{IEEEtran}
\end{document}